\documentclass[nature,preprint,superscriptaddress,amssymb,amsmath]{revtex4-1}
\usepackage{graphicx}
\usepackage{dcolumn}
\usepackage{bm, color}
\usepackage[none]{hyphenat} 
\usepackage{amsmath}
\usepackage{lipsum}

\begin{document}

\title{Giant anomalous Hall and Nernst effect in magnetic cubic Heusler compounds}
\author{Jonathan Noky}
\affiliation{Max Planck Institute for Chemical Physics of Solids, D-01187 Dresden, Germany}
\author{Yang Zhang}
\affiliation{Max Planck Institute for Chemical Physics of Solids, D-01187 Dresden, Germany}
\affiliation{Massachusetts Institute of Technology, Cambridge, MA 02139, USA}
\author{Claudia Felser} 
\affiliation{Max Planck Institute for Chemical Physics of Solids, D-01187 Dresden, Germany}
\author{Yan Sun} 
\email{ysun@cpfs.mpg.de}
\affiliation{Max Planck Institute for Chemical Physics of Solids, D-01187 Dresden, Germany}

\date{\today}

\begin{abstract}

The interplay of magnetism and topology opens up the possibility for exotic linear response effects, such as the anomalous Hall effect and the anomalous Nernst effect, which can be strongly enhanced by designing a strong Berry curvature in the electronic structure. It is even possible to utilize this to create a quantum anomalous Hall state at high temperatures by reducing the dimensionality. Magnetic Heusler compounds are a promising class of materials for this purpose because they grow in thin films, have a high Curie
temperature, and their electronic structure hosts strong topological features. Here, we provide
a comprehensive study of the intrinsic anomalous transport for magnetic cubic full
Heusler compounds and we illustrate that several Heusler compounds outperform the best
so far reported materials. The results reveal the importance of symmetries, especially mirror planes, in
combination with magnetism for giant anomalous Hall and Nernst effects, which should
be valid in general for linear responses (spin Hall effect, spin orbital torque, etc.)
dominated by intrinsic contributions.

\end{abstract}
\maketitle
\section{Introduction}

In recent years a huge effort has been put in identifying topological phases of matter in real materials, culminating in a way to classify all materials via elementary band representation theory~\cite{bradlyn2017topological,cano2018topology,cano2018building,vergniory2019complete}. However, until now this is done specifically for time-reversal symmetric, i. e. non-magnetic materials. For the investigation of magnetic materials the only possibility is to perform systematic calculations for each single compound which has been done for selected material classes~\cite{PhysRevMaterials.3.024410,PhysRevMaterials.3.024005,Sanvitoe1602241}. The restriction to time-reversal symmetric systems excludes not only a large number of compounds but also all kinds of properties that require broken time-reversal symmetry. Specifically, in magnetic materials there exist linear response effects that are not possible in time-reversal symmetric systems: There is the anomalous Hall effect (AHE)~\cite{Pugh_1953,nagaosa2010anomalous}, which describes the transverse voltage drop that results from an applied longitudinal current, and the anomalous Nernst effect (ANE)~\cite{Nerst_1887,XiaoDi_2006}, which is analogous to the AHE but with a longitudinal temperature gradient instead of a current. 

Furthermore, it is proposed that a system with a large AHE in the bulk phase can undergo a transition into a quantum anomalous Hall (QAH)~\cite{haldane1988qahe,Chang2013} state when preparing it as a thin film~\cite{xu2011chern,liu2017quantum,muechler2017realization}. Here, the topological features that cause the large AHE, e.g. Weyl points or nodal lines, are gapped in the thin film limit, leading to an inverted band gap with a QAH effect. This can even be realized at high temperatures when the film has an appropriate Curie temperature and can be utilized for applications in topological quantum computing~\cite{sarma2006topological}.

A promising class of candidate materials are the magnetic Heusler compounds~\cite{heusler1903ubermagnetische}. They are
a versatile group of materials that are easily tunable for many different properties~\cite{chadov2010,wollmann2017,manna2017colossal,Manna2018}.
Additionally, they have large Curie temperatures and it is possible to grow them
as thin films~\cite{glas2013anomalous,hu2018anomalous,reichlova2018large,markou2019thickness}.
Heusler compounds host both magnetism and topological band structures, leading to a
very large Berry curvature (BC). A strong AHE and the highest so far reported ANE were found in a Heusler system,
Co$_2$MnGa~\cite{Belopolski1278,ludbrook2017perpendicular,Guin2018,Sakai2018,reichlova2018large}. 
This renders them as a promising platform for
investigating the anomalous linear response effects like the AHE, the anomalous Nernst effect~\cite{Nerst_1887,XiaoDi_2006} (ANE), and the magneto-optic Kerr effect~\cite{kerr1877,PhysRevB.75.214416,kahn1969ultraviolet} (MOKE), all of these have their intrinsic contributions enhanced due to the BC~\cite{nagaosa2010anomalous,XiaoDi_2006,Thouless_1982,Xiao2010}.

Studying these effects gives insights into
the electrical, thermoelectrical, and magneto-optical properties of magnetic materials~\cite{avron2003topological} and can also be utilized in applications, e.g. for data storage, data transfer, and thermoelectric
power generation. Therefore, both the understanding of the underlying mechanisms and
the search for materials with strong linear responses in the effects described above has
attracted extensive interest in both fundamental physics and material science. However, the plethora of possible compounds also makes it
challenging to figure out the most promising ones. Therefore, simple guidelines for
estimates of the linear responses are required to open up new possibilities in the search for
interesting materials with large linear response effects.

\section{Results and discussion}

\subsection{Guiding principles for large AHE and ANE}

\begin{figure}[htb]
\centering
\includegraphics[width=0.3\textwidth]{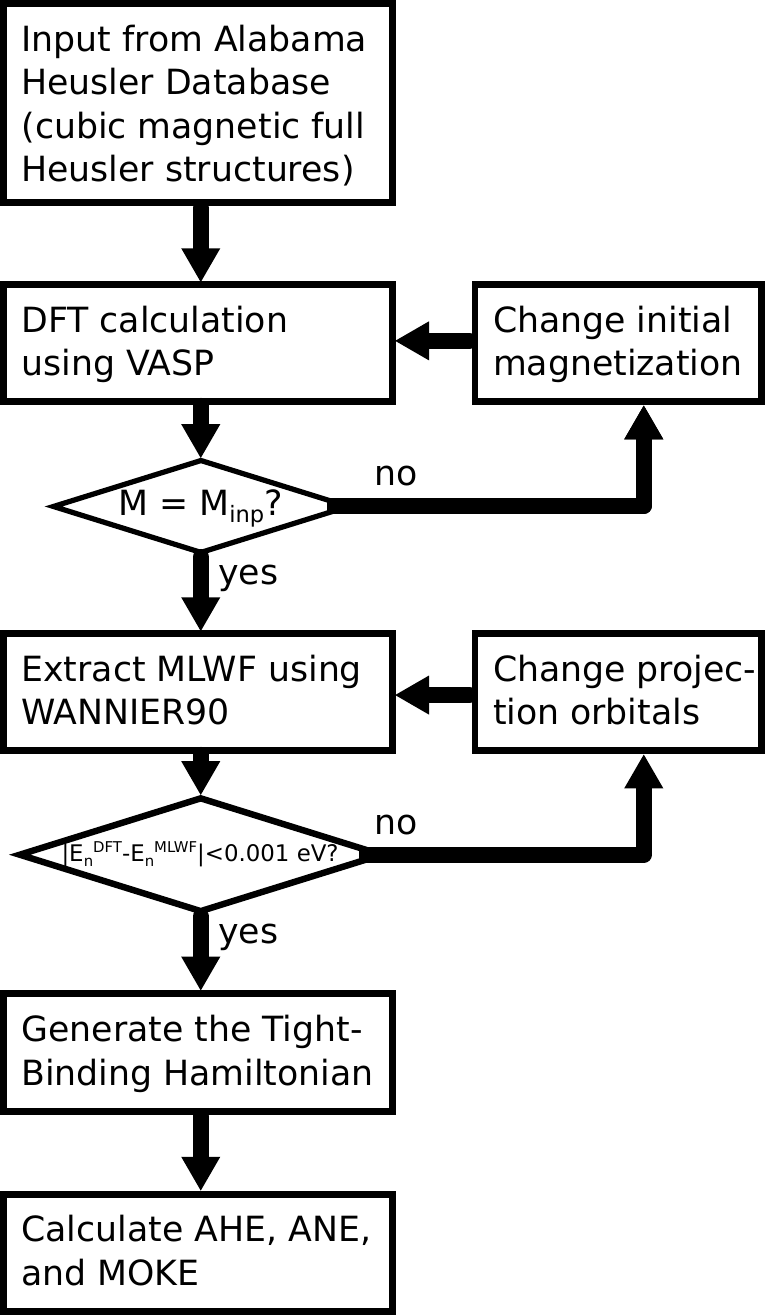}
   \caption{Workflow for the analysis of the materials. All stable, cubic, and magnetic full Heusler structures from the Alabama Heusler Database~\cite{alabama} are taken as a starting input.}
\label{fig:0}
\end{figure}

In this work we theoretically investigate all cubic and stable magnetic full Heusler compounds from the Heusler database of the University of Alabama~\cite{alabama}. This is done by applying the workflow shown in Figure~\ref{fig:0} to all compounds which satisfy the above mentioned criteria. We link the resulting anomalous Hall conductivity (AHC), anomalous Nernst conductivity (ANC), and Kerr angle to structural and electronic properties of the materials, such as space group (SG), lattice constant, magnetic moment, number of valence electrons, and density of states at the Fermi level.

\begin{figure}[htb]
\centering
\includegraphics[width=0.48\textwidth]{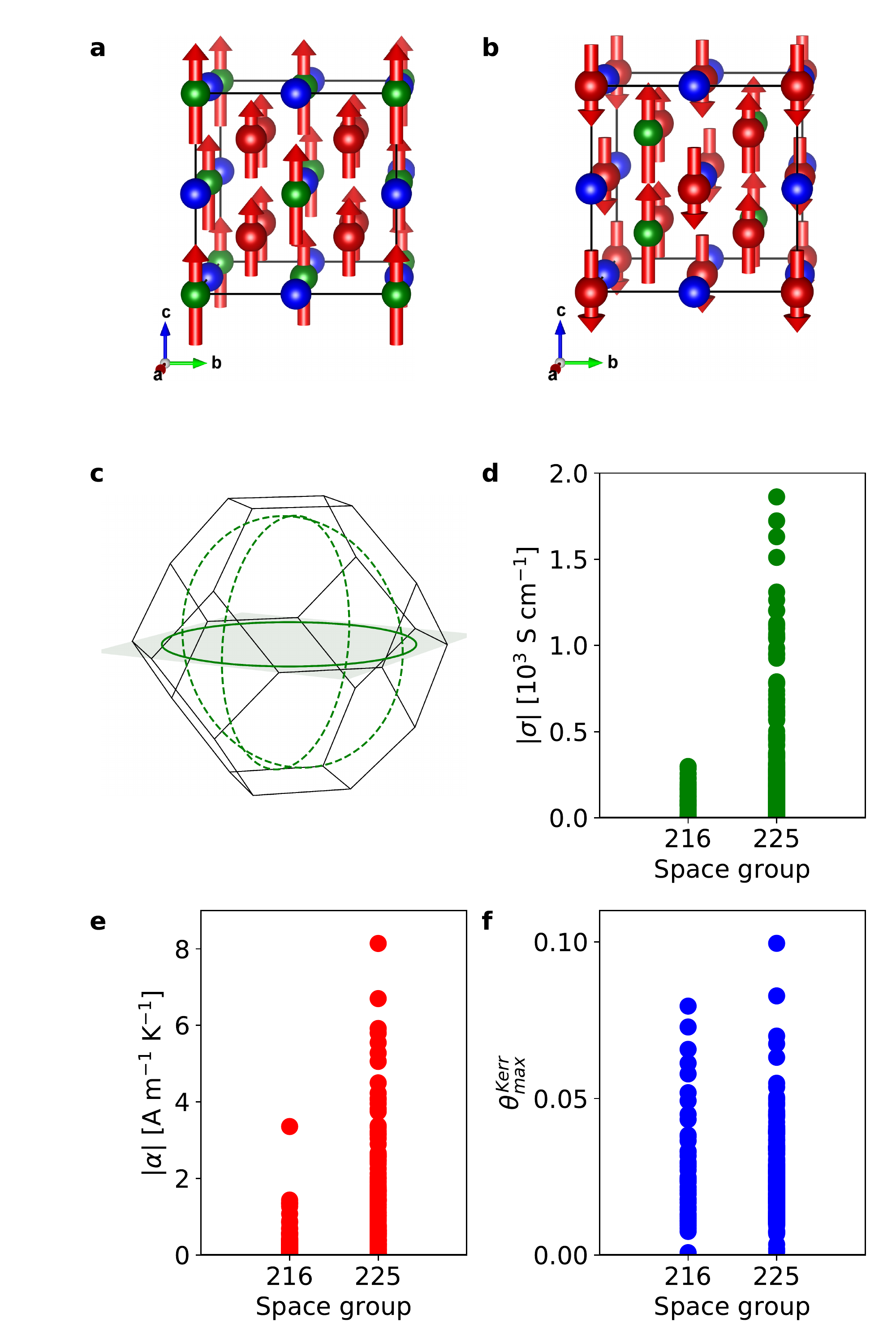}
   \caption{Comparison of the different space groups. (a), (b) Crystal structure of the regular (space group 225) and inverse (space group 216) Heusler structure, respectively. (c) Brillouin zone of the cubic Heusler compounds. The green lines represent the nodal lines which are enforced by symmetry in the regular compounds. When choosing the magnetic moments aligned along (001), only the horizontal mirror plane (green) is preserved. Consequently, the dashed nodal lines gap out. (d) Anomalous Hall conductivity (AHC), $\sigma$, (e) anomalous Nernst conductivity (ANC), $\alpha$, and (f) maximum Kerr angle, $\theta_{max}^{Kerr}$, in a range from 1 eV to 4 eV in dependence of the space group. Space group 225 shows the larger values for all three responses.}
\label{fig:1}
\end{figure}

To reveal general rules for the linear response in cubic Heusler compounds, we analyzed these results carefully for underlying concepts and correlations. A distinct difference is present between the two space groups. In Figure~\ref{fig:1}(a) and (b) the crystal structure of a regular (SG 225) and an inverse (SG 216) cubic Heusler compound are shown, respectively. As the main difference between the two groups, SG 225 hosts three mirror planes at $x=0$, $y=0$, and $z=0$ that are not present in SG 216 due to the reduced symmetry. Comparing now the values for the AHC, ANC, and Kerr angle of both space groups, we find larger values of all three properties for the regular structure (SG 225), as shown in Figure~\ref{fig:1}(d)-(f).

\begin{figure}[htb]
\centering
\includegraphics[width=0.4\textwidth]{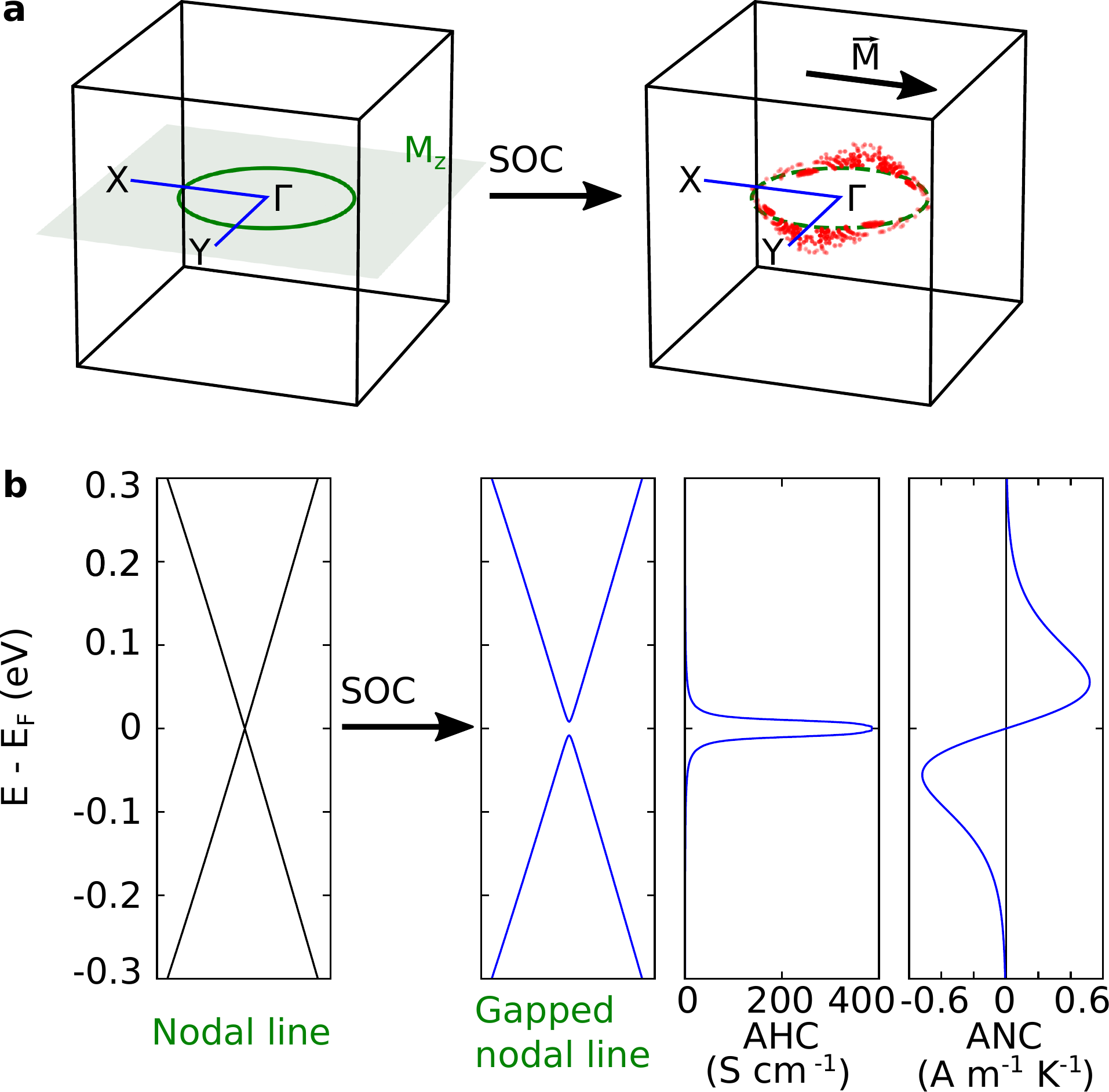}
   \caption{Nodal line model system. (a) Left panel: System without spin-orbit coupling. The mirror symmetry of the model protects a nodal line in the $k_x$-$k_y$ plane. Right panel: Introducing spin-orbit coupling breaks the mirror symmetry and gaps out the nodal line, inducing large Berry curvature around the former nodal line. (b) Band structures of the systems without and with spin-orbit coupling, respectively. The gapped nodal line creates large anomalous Hall and Nernst effects.}
\label{fig:mod}
\end{figure}

For an understanding of the underlying mechanism of this difference, the presence of mirror planes plays a crucial role. To elaborate on the influence of this specific symmetry, we analyze a simple four-band model proposed by Rauch \textit{et al.}~\cite{PhysRevB.96.235103}
\begin{align}
        H=(m-6M&+2(\cos{k_x}+\cos{k_y}+\cos{k_z}))\tau_z\otimes\sigma_0+\nonumber\\ &+B\tau_z\otimes\sigma_z+c\sin{k_z}\tau_x\otimes\sigma_z\nonumber\\&+c\sin{k_x}\tau_x\otimes\sigma_x+c\sin{k_y}\tau_x\otimes\sigma_y.
\end{align}
For $m=M=c=1$ eV and $B=2$ eV this model possesses a mirror in the $k_x$-$k_y$-plane, that hosts a nodal line (NL) at the Fermi level protected from this symmetry (Figure~\ref{fig:mod}(a) left panel). We now introduce a term for spin-orbit coupling (SOC), that can be interpreted as a consequence of a magnetization along the $x$-direction~\cite{noky2018characterization,noky2018large}
\begin{equation}
        H^{SOC}=\lambda c(\sin{k_x}\tau_x\otimes\sigma_x+\sin{k_z}\tau_x\otimes\sigma_y)\text{ with }\lambda=0.01
\end{equation}
Because a magnetization along the $x$-direction in combination with SOC breaks the mirror symmetry in the $k_x$-$k_y$ plane, the NL is no longer protected and gets gapped. The created gap is an inverted band gap that induces strong BC into the band structure. The distribution of the BC in the Brillouin zone is focused only around the former NL, as shown in Figure~\ref{fig:mod}(a) right panel. Consequently, the now gapped system exhibits a large AHE and ANE around the Fermi level.

From this model the larger linear response values in the materials of SG 225 can be understood. Because SG 225 hosts three mirror planes perpendicular to each other, there are three NLs protected by these symmetries. However, all the investigated compounds are magnetic, therefore depending on the magnetization direction the mirror symmetries are broken 
and the NLs are no longer protected (see Figure~\ref{fig:1}(c) for a magnetization along (001)). Consequently, the respective NLs gap out and induce large BC into the band structure~\cite{kim2018large,noky2018large,noky2018characterization}. This effect leads to the enhanced values in AHC, ANC, and the Kerr angle, as they are closely related to the BC (see \textit{Methods section} and Supplementary Figure S1).

\begin{figure}[htb]
\centering
\includegraphics[width=0.48\textwidth]{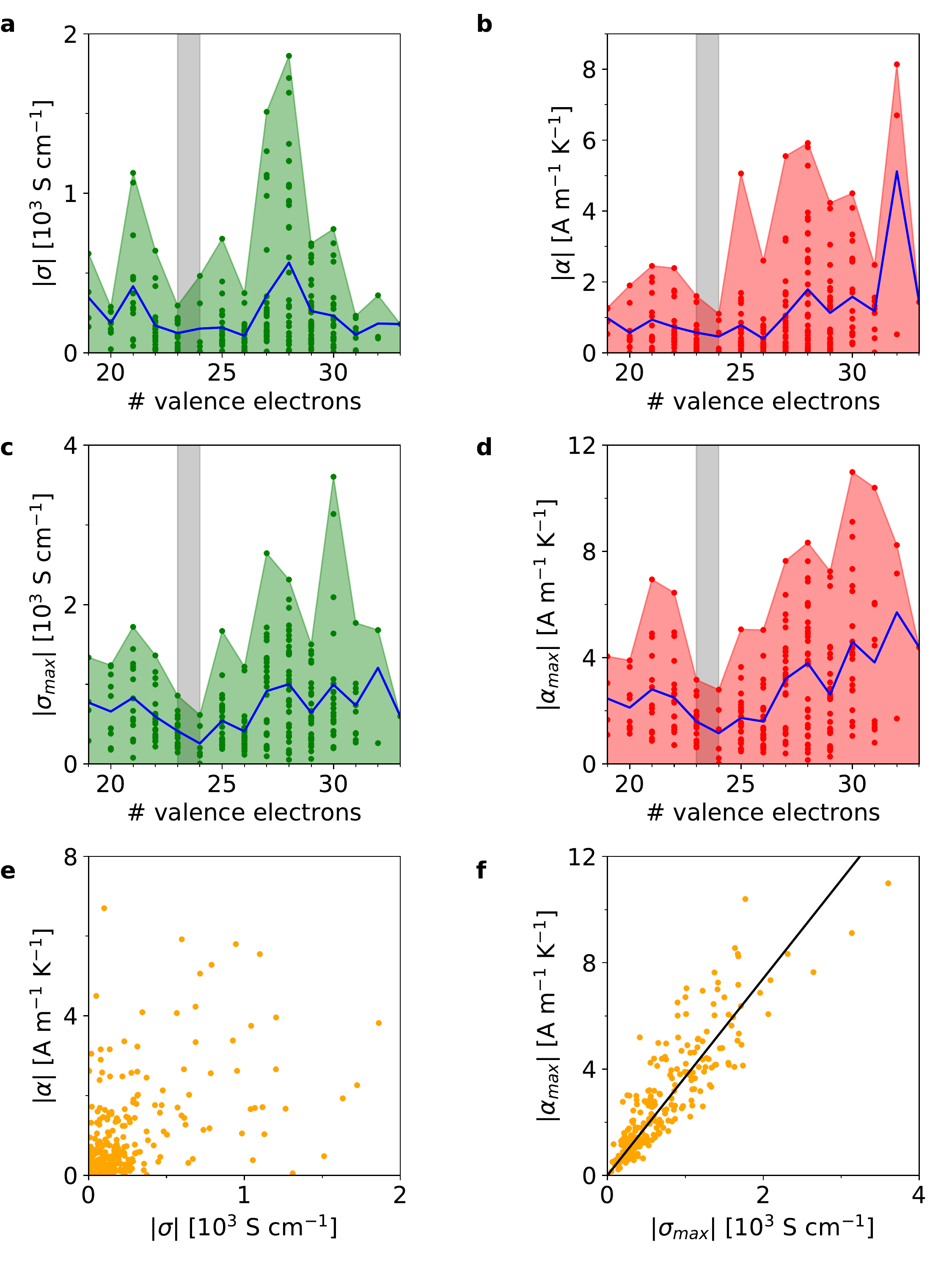}
   \caption{Connections between the anomalous Hall conductivity (AHC) and the anomalous Nernst conductivity (ANC). (a) AHC, $\sigma$, and (b) ANC, $\alpha$, in dependence of the number of valence electrons. (c) Maximum of $\sigma$ and (d) $\alpha$ in a window of 250 meV around $E_F$ in dependence of the number of valence electrons. The green/red area connects the maximum values for each valence electron count. The blue line shows the average value. The grey area is a guide to the eye for the minimum area at 23 to 24 valence electrons between the two peaks at 21 and 28 valence electrons. (e) $\alpha$ in dependence of $\sigma$. No correlation is visible. (f) Maximum of $\alpha$ in dependence of the maximum of $\sigma$. A clear linear connection can be seen.}
\label{fig:2}
\end{figure}

It is also interesting to investigate the connection between the number of valence electrons and the AHC/ANC. Figure~\ref{fig:2}(a) reveals a double peak structure in the AHC with a small maximum located at 21 valence electrons and a larger maximum at 28. This can be seen for both maximum values (light green area) and average values (blue line). A similar behavior of a double peak structure with a minimum around 23 to 24 valence electrons (grey area) exists for the ANC in Figure~\ref{fig:2}(b). Due to the fact that both AHE and ANE are very sensitive to the Fermi level $E_F$ we also investigate the maximum AHC and ANC in a range of 250 meV around $E_F$. Scanning a range of possible Fermi energies is especially interesting because Heusler compounds are easy to dope and thus the Fermi level can be controlled by changing the compositions slightly. Figure~\ref{fig:2}(c) and (d) show the dependence of maximum AHC and ANC, respectively, on the number of valence electrons. Also here the double peak structure with the same minimum position can be identified. 

This behavior can be understood when taking the energetic position of the BC inducing features into account. Because the valence electron count can be related to the filling of the band structure, certain electron numbers move $E_F$ close to the topological features while at other fillings there is no strong BC at $E_F$. A valence electron count of 21 or 28 corresponds to a filling level with the topological features close to $E_F$, which leads in consequence to large AHC and ANC.

Another interesting connection to investigate is between the different responses. Looking at the dependence of the ANC on the AHC in Figure~\ref{fig:2}(e) it can be seen that the two properties are independent of each other, i.~e. a large AHC not necessarily causes a large ANC. This shows the importance of treating the two mechanisms independently even if they are both linked to the BC~\cite{noky2018characterization}. However, a different picture arises when taking the maximum value of both AHC and ANC in a range of 250 meV around $E_F$ into account, as shown in Figure~\ref{fig:2}(f). Here, a linear connection between the absolute values is visible. This can be understood from the fact that both effects rely on a large BC in the band structure. Due to the different mechanisms of the two effects (see \textit{methods section}) the energetic distribution of the contributions is different but nonetheless the presence of large BC increases both of these values~\cite{noky2018large,noky2018characterization}.

In conclusion, the statistic analysis we performed in the previous paragraphs shows, that the symmetry of the compound is very important to find materials with large linear responses. We find, that mirror symmetries in combination with magnetism are crucial for getting strong effects. Apart from that the most important influence is the exact location of the BC in the band structure with respect to the Fermi level. For full Heusler compounds our calculations reveal that this is the case for a number of valence electrons around 21 and 28. We do not find such clear correlations for the Kerr angle because it is sensitive not only to the Fermi level position but to the transition energy $\hbar\omega$.

In addition to these general results, the calculations for the single materials show very appealing properties. In particular, some of the calculated values are in range of or even exceed the highest so far reported values.

\subsection{Selected materials with large AHE and ANE}

In the following we present selected results of the calculated compounds. The full results for the 255 compounds are shown in the Supplementary Information. Table~\ref{tab:1} shows materials with a large AHC and/or ANC.

Looking at the AHC we find that the compounds Co$_2$MnAl, Rh$_2$MnAl, and Rh$_2$MnGa have values almost as large as Fe~\cite{miyasato2007crossover}, which is the highest value ever reported. This is especially interesting because the first two mentioned compounds have already been synthesized~\cite{WEBSTER19711221,masumoto1972new}. For the ANC the largest reported experimental value so far is 6 A m$^{-1}$ K$^{-1}$ in Co$_2$MnGa~\cite{Sakai2018,Guin2018}. It is important to note, that this value is not reached at the charge neutral point but with some doping~\cite{Guin2018}. In the Heusler compounds we find a similar value for Rh$_2$NiSi and an even larger value for Rh$_2$NiSn, which has already been synthesized~\cite{SUITS1976423}. 

However, it is important to also include the energy dependence of the AHC and ANC to account for energetic shifts away from the charge neutral point. Therefore, we additionally show the largest possible value in a range of 250 meV around $E_F$ in Table~\ref{tab:1}. This is, because real materials will always have some vacancies or defects which lead to small doping effects and consequently to a shift in the Fermi level. Furthermore, Heusler compounds can also be doped in a controlled way to engineer a compound with the desired Fermi level position. Taking the possibility of doping into account, all compounds shown in Table~\ref{tab:1} have a very large ANC, with most of them even exceeding the current record value. The largest values that we find are 10.99 A m$^{-1}$ K$^{-1}$ in Co$_2$FeSn, 9.11 A m$^{-1}$ K$^{-1}$ in Co$_2$FeGe, and 8.33 A m$^{-1}$ K$^{-1}$ in Rh$_2$MnGa.

\begin{figure*}[htb]
\centering
\includegraphics[width=0.96\textwidth]{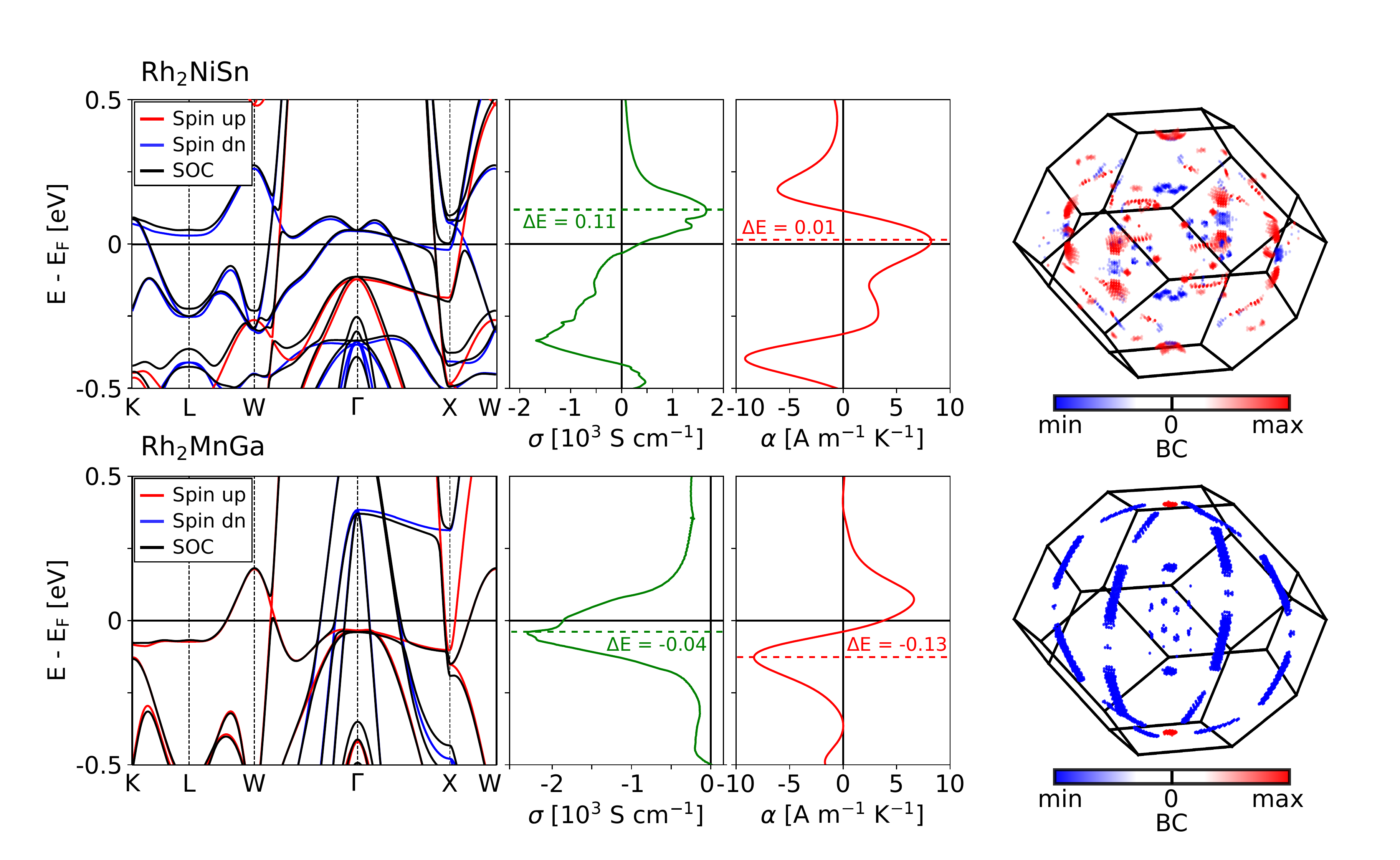}
   \caption{Band structure with and without spin-orbit coupling, anomalous Hall conductivity (AHC), $\sigma$, anomalous Nernst conductivity (ANC), $\alpha$, and Berry curvature distribution in the Brillouin zone for Rh$_2$NiSn and Rh$_2$MnGa. Both AHC and ANC are strongly dependent on the position of the Fermi level. Note the scaling factor in the AHC. The nodal lines in Rh$_2$MnGa are clearly visible.}
\label{fig:4}
\end{figure*}

These results show that the changes in $E_F$ can have a large influence on the anomalous transport coefficients, especially on the ANC. To illustrate this more in detail, Figure~\ref{fig:4} shows two example materials from Table~\ref{tab:1} with the energy dependent AHC and ANC. In both band structures in Figure~\ref{fig:4} the effect shown in the tight-binding model of the NL can be seen: For the calculation without SOC there is a band crossing along the $W$-$\Gamma$ line, which gets gapped by the introduction of SOC. It is important to note, that the corresponding NLs also have a dispersion in energy, which is the reason that the peak in the AHC does not coincide with the crossing along the $W$-$\Gamma$ line.

For the AHC a small shift in the Fermi level can greatly enhance or decrease the value, as it can be seen for Rh$_2$NiSn (Figure~\ref{fig:4} upper panel): Here the AHC is 1682 S cm$^{-1}$ for $E-E_F=-110$ meV but only 360 S cm$^{-1}$ at $E_F$. However, for the ANC these changes are even stronger. This can be seen in Rh$_2$MnGa (Figure~\ref{fig:4} lower panel), where a change from $E_F$ to $E-E_F=-130$ meV leads to a change in the ANC from 3.82  A m$^{-1}$ K$^{-1}$ to -8.33 A m$^{-1}$ K$^{-1}$, including a sign change. 

Therefore, it is very important for comparison between experimental and calculated results to take the exact position of the Fermi level into account. The energy dependence of the AHC and ANC for all calculated materials is given in the Supplementary Information.

There can be different sources of the BC in a band structure. Mainly, it can come from Weyl points (WP) that act as monopoles for the BC or from inverted band gaps, that are created from a gapped out NL in these compounds, when a mirror symmetry is broken via magnetization. While WPs are present in most of the Heusler compounds in this work, NLs are only possible in SG 225. Here, both mechanisms are present. To find out the dominating effect, one can look into the BC distribution in the Brillouin zone. In the case of WPs the distribution is mainly point-like, as it can be seen for Rh$_2$NiSn in Figure~\ref{fig:4}. For the NL case, also the BC forms a line following the former NL like in Rh$_2$MnGa in Figure~\ref{fig:4}. For the high-performance compounds in Tab.~\ref{tab:1} the last column gives the dominating mechanism.

\subsection{Summary}

In summary, we have investigated the AHE, ANE, and MOKE of all the stable magnetic cubic full Heusler compounds given in the Heusler database of the university of Alabama\cite{alabama}. From evaluating the linear response values of the different space groups we find that the symmetries are very important. To achieve a large AHC and ANC the presence of mirror symmetries turns out to be crucial, which leads in combination with magnetism to a large BC in the band structure and consequently to large linear response coefficients. The existance of mirror planes should also be important for other linear response effects that are dominated by intrinsic contributions such as the spin Hall effect, spin Nernst effect, and spin-orbit torgue.

Apart from that the linear response is mostly influenced by the exact position of the BC in the band structure with respect to the Fermi level. For Heusler compounds, this can be linked to the number of valence electrons with two sweet spots at 21 and 28 electrons per unit cell. Some of the materials show a very large AHC and/or ANC, which is close to or even exceeds the highest values that are reported so far. It is important to note, that both AHC and ANC are strongly dependent on the position of the Fermi level and therefore on the doping of the investigated material. This always has to be taken into account when comparing experimental and theoretical results.

This work shows the versatility and usability of Heusler compounds to achieve large linear response values and illustrates the importance of mirror symmetries. It proposes a comprehensive list of their properties including large values for the AHC and new record values for the ANC and thus paves the way for new high-performance compounds.

\section{Methods}

For our investigation we start with the structural data given by the Heusler Database of the University of Alabama~\cite{alabama} where we choose all cubic compounds that are given as stable. The compounds without a magnetization are set aside because for the investigated linear responses only systems without time-reversal symmetry are interesting. This results in 255 materials which satisfy these conditions.

For the chosen compounds we take the lattice constant and the space group as inputs for the density-functional theory (DFT) calculation. For this we employ the package VASP~\cite{kresse1996} with pseudopotentials, plane waves and the generalized-gradient approximation (GGA)~\cite{perdew1996} for the exchange-correlation potential. For the self-consistent calculations a $k$ mesh of $13\times 13\times 13$ points was used. The DFT calculation is done with different starting values for the magnetic moments until the total magnetization given in the database is reproduced. For all calculations the magnetic moments were chosen to be parallel to the (001) direction. In the following we create Wannier functions via the package Wannier90~\cite{Mostofi2008} with a projection of the Bloch states to the atomic orbitals. The wannierization is repeated with different parameters until the average energy difference between the Wannier functions and the DFT wave functions along a given path in \textit{k} space is less than 1 meV in a range from $E_F-4$ eV to $E_F+4$ eV. Then we extract Tight-Binding parameters and use this Hamiltonian, $H$, to calculate the Berry curvature (BC), $\Omega$, via the Kubo formalism~\cite{Thouless_1982,Xiao2010,nagaosa2010anomalous}
\begin{equation}
  \Omega_{ij}^n=\sum_{m\ne n} \frac{\langle n|\frac{\partial H}{\partial k_i}|m\rangle \langle m|\frac{\partial H}{\partial k_j}|n\rangle - (i \leftrightarrow j)}{(E_n-E_m)^2},
\end{equation}
with $|n\rangle$ and $E_n$ being the eigenstates and -energies of $H$, respectively.

We evaluate the anomalous Hall conductivity (AHC), $\sigma$, from the BC as
\begin{equation}
 \label{eq:ahc}
 \sigma_{ij}=\frac{e^2}{\hbar} \sum_n\int \frac{d^3k}{(2\pi)^3}\Omega_{ij}^nf_n
\end{equation}
and the anomalous Nernst conductivity (ANC), $\alpha$, as proposed by Xiao et al. \cite{Xiao2010, XiaoDi_2006}
\begin{align}
  \label{eq:anc}
  \alpha_{ij}=-\frac{1}{T} \frac{e}{\hbar} \sum_n \int \frac{d^3k}{(2\pi)^3} \Omega_{ij}^{n}[(E_{n}-E_F)f_{n}+\nonumber \\
  +k_BT\ln{(1+\exp{\frac{E_{n}-E_F}{-k_BT}})}].
\end{align}
Here, $T$ is the actual temperature, $f_n$ is the Fermi distribution, and $E_F$ is the Fermi level.

Furthermore, we calculate the optical Hall conductivity, $\sigma(\omega)$, using the Kubo formalism~\cite{PhysRevLett.92.037204,ebert1996magneto}:
\begin{align}
\label{eq:kerr}
 \sigma_{ij}(\omega)=\frac{e^2}{\hbar}\sum_{n\neq m}\int \frac{d^3k}{(2\pi)^3} (f_{n,k}-f_{m,k})\times \nonumber \\
 \times\frac{\langle n|\frac{\partial H}{\partial k_i}|m\rangle \langle m|\frac{\partial H}{\partial k_j}|n\rangle - (i \leftrightarrow j)}{(E_n-E_m)^2-(\hbar\omega+i\delta)^2},
\end{align}
where $\omega$ is the transition energy and $\delta$ has a positive infinitesimal value. From this we evaluate the Kerr angle, $\theta_K$, as~\cite{PhysRevB.75.214416,kahn1969ultraviolet}
\begin{equation}
 \theta_K+i\eta_K=\frac{-\sigma_{xy}}{\sigma_{xx}\sqrt{1+\frac{4\pi i}{\omega}\sigma_{xx}}}.
\end{equation}

\begin{table*}
\scalebox{0.9}{%
\begin{tabular}{|c|c|c|c|c|c|c|c|c|} \hline
Material & SG & $a_0$ & $\mu$ & AHC & AHC$_{max}$ ($\Delta E$/$\Delta n$) & ANC & ANC$_{max}$ ($\Delta E$/$\Delta n$)  & Main BC\\ 
  &  & (\AA)  &($\mu_B$/f.u.)  & (S cm$^{-1}$) & (S cm$^{-1}$) & (A m$^{-1}$ K$^{-1}$) & (A m$^{-1}$ K$^{-1}$) & source  \\ \hline 
Co$_2$CrAl & 225 & 5.7 & 3.0 &  -313 & -1089 (-0.04/-0.24) & 3.23 & 3.23 (0.0/-0.03) & WP\\ \hline
Co$_2$MnAl & 225 & 5.7 & 4.04 & -1631 & -1739 (-0.01/-0.02) & 1.93 & 4.13 (0.04/0.05) & NL\\ \hline
Co$_2$MnGa  & 225 & 5.72 & 4.11 & -1310 & -1473 (0.05/0.1) & -0.05 & 4.79 (0.09/0.17) & NL \\ \hline
Co$_2$FeSi & 225 & 5.63 & 5.4 & -275 & 2092 (0.12/0.8) & 2.57 & 7.34 (0.08/0.49) & WP\\ \hline
Co$_2$FeGe & 225 & 5.74 & 5.57 & -78 & 3136 (0.14/0.84) & 3.16 & 9.11 (0.08/0.43) & NL\\ \hline
Co$_2$FeSn & 225 & 5.99 & 5.6 & 49 & 3602 (0.12/0.85) & 4.5 & 10.99 (0.07/0.4) & NL\\ \hline
Fe$_2$MnP & 225 & 5.55 & 4.0 & -1202 & -1373 (-0.04/-0.13) & 2.66 & 7.63 (0.12/0.37) & NL\\ \hline
Fe$_2$MnAs & 225 & 5.7 & 4.02 & -1043 & -1413 (-0.06/-0.2) & 3.75 & 7.0 (0.1/0.31) & NL\\ \hline
Fe$_2$MnSb & 225 & 5.95 & 4.11 & -1203 & -1374 (-0.06/-0.31) & 3.96 & -6.02 (-0.12/-0.67) & NL\\ \hline
Rh$_2$MnAl & 225 & 6.04 & 4.1 & -1723 & -2064 (-0.05/-0.13) & 2.26 & -6.06 (-0.12/-0.39) & NL\\ \hline
Rh$_2$MnGa & 225 & 6.06 & 4.13 & -1862 & -2313 (-0.04/-0.13) & 3.82 & -8.33 (-0.13/-0.45) & NL\\ \hline
Rh$_2$FeIn & 225 & 6.27 & 4.2 & -18 & 1270 (0.12/0.63) & 3.05 & 4.42 (0.07/0.33) & WP\\ \hline
Rh$_2$CoAl & 225 & 5.98 & 3.03 & 345 & 1005 (0.05/0.26) & 4.09 & -4.2 (0.12/0.57) & NL\\ \hline
Rh$_2$NiSi & 225 & 5.89 & 0.98 & 100 & 1678 (0.14/0.6) & 6.7 & 7.17 (0.04/0.14) & WP\\ \hline
Rh$_2$NiSn & 225 & 6.21 & 1.0 & 360 & 1680 (0.11/0.55) & 8.14 & 8.24 (0.01/0.04) & WP\\ \hline
Ru$_2$MnP & 225 & 5.91 & 3.95 & -926 & -1159 (0.03/0.08) & -3.38 & -5.12 (-0.05/-0.15) & NL\\ \hline
Ru$_2$FeP & 225 & 5.9 & 4.13 & -686 & -1300 (0.04/0.17) & -4.23 & -4.37 (-0.01/-0.05) & WP\\ \hline
Ru$_2$FeAs & 225 & 6.02 & 4.23 & -566 & -1378 (0.08/0.32) & -4.07 & 4.15 (0.16/0.74) & WP\\ \hline
Ru$_2$CoGe & 225 & 5.96 & 1.99 & -138 & -1500 (0.15/0.91) & -2.48 & -6.7 (0.11/0.68) & WP\\ \hline
Ru$_2$CoP & 225 & 5.84 & 2.18 & -776 & -899 (-0.03/-0.14) & 1.18 & -6.5 (-0.12/-0.62) & NL\\ \hline
\end{tabular}}
\caption{Results for selected materials with large anomalous Hall conductiviy (AHC) and/or  anomalous Nernst conductivity (ANC). The full results are shown in the Supplementary Information. Listed are: space group (SG), theoretical lattice constant ($a_0$), theoretical magnetic moment ($\mu$), AHC, maximum AHC, ANC, maximum ANC. The maximum values are obtained in an energy window of 250 meV around the Fermi level. $\Delta E$ gives the energy distance in meV and $\Delta n$ the electron difference of the maximum values with regard to the Fermi level. The last column gives the main source of Berry curvature (BC) as either Weyl points (WP) or gapped nodal lines (NL). Very large values for both AHC and ANC are possible.}
\label{tab:1}
\end{table*}

\section{Data availability}

All data generated and/or analyzed during this study are included in this article and its Supplementary Information file. The data are available from the corresponding author upon reasonable request.

\section{Acknowledgements}

We are grateful to J. Gooth for insightful discussions. This work was financially supported by the ERC Advanced Grant No. 291472 'Idea Heusler', ERC Advanced Grant No. 742068 'TOPMAT', and 'ASPIN' (EU H2020 FET Open Grant No. 766566).

\section{Author Information}

\subsection{Affiliations}

\textit{Max Planck Institute for Chemical Physics of Solids, Dresden, Germany} \\
Jonathan Noky, Yang Zhang, Claudia Felser, Yan Sun\\ \\
\textit{Massachusetts Institute of Technology, Cambridge, USA} \\Yang Zhang

\subsection{Contributions}

J.N. carried out all calculations except the MOKE, analysed the results, and wrote the manuscript. Y.Z. carried out the MOKE calculations. Y.S. supervised J.N. and Y.Z. All authors discussed the results and commented on the manuscript.

\subsection{Competing interests}

The authors declare no competing interests.

\subsection{Corresponding author}

Correspondence to Yan Sun (ysun@cpfs.mpg.de).

\bibliographystyle{naturemag}
\bibliography{Heusler2}

\begin{thebibliography}{10}
\expandafter\ifx\csname url\endcsname\relax
  \def\url#1{\texttt{#1}}\fi
\expandafter\ifx\csname urlprefix\endcsname\relax\def\urlprefix{URL }\fi
\providecommand{\bibinfo}[2]{#2}
\providecommand{\eprint}[2][]{\url{#2}}

\bibitem{bradlyn2017topological}
\bibinfo{author}{Bradlyn, B.} \emph{et~al.}
\newblock \bibinfo{title}{{Topological quantum chemistry}}.
\newblock \emph{\bibinfo{journal}{Nature}} \textbf{\bibinfo{volume}{547}},
  \bibinfo{pages}{298--305} (\bibinfo{year}{2017}).

\bibitem{cano2018topology}
\bibinfo{author}{Cano, J.} \emph{et~al.}
\newblock \bibinfo{title}{{Topology of Disconnected Elementary Band
  Representations}}.
\newblock \emph{\bibinfo{journal}{Physical Review Letters}}
  \textbf{\bibinfo{volume}{120}}, \bibinfo{pages}{266401}
  (\bibinfo{year}{2018}).

\bibitem{cano2018building}
\bibinfo{author}{Cano, J.} \emph{et~al.}
\newblock \bibinfo{title}{{Building blocks of topological quantum chemistry:
  Elementary band representations}}.
\newblock \emph{\bibinfo{journal}{Physical Review B}}
  \textbf{\bibinfo{volume}{97}}, \bibinfo{pages}{35139} (\bibinfo{year}{2018}).

\bibitem{vergniory2019complete}
\bibinfo{author}{Vergniory, M.~G.} \emph{et~al.}
\newblock \bibinfo{title}{{A complete catalogue of high-quality topological
  materials}}.
\newblock \emph{\bibinfo{journal}{Nature}} \textbf{\bibinfo{volume}{566}},
  \bibinfo{pages}{480--485} (\bibinfo{year}{2019}).

\bibitem{PhysRevMaterials.3.024410}
\bibinfo{author}{Gao, Q.}, \bibinfo{author}{Opahle, I.} \&
  \bibinfo{author}{Zhang, H.}
\newblock \bibinfo{title}{High-throughput screening for spin-gapless
  semiconductors in quaternary heusler compounds}.
\newblock \emph{\bibinfo{journal}{Phys. Rev. Materials}}
  \textbf{\bibinfo{volume}{3}}, \bibinfo{pages}{024410} (\bibinfo{year}{2019}).

\bibitem{PhysRevMaterials.3.024005}
\bibinfo{author}{Olsen, T.} \emph{et~al.}
\newblock \bibinfo{title}{Discovering two-dimensional topological insulators
  from high-throughput computations}.
\newblock \emph{\bibinfo{journal}{Phys. Rev. Materials}}
  \textbf{\bibinfo{volume}{3}}, \bibinfo{pages}{024005} (\bibinfo{year}{2019}).

\bibitem{Sanvitoe1602241}
\bibinfo{author}{Sanvito, S.} \emph{et~al.}
\newblock \bibinfo{title}{Accelerated discovery of new magnets in the heusler
  alloy family}.
\newblock \emph{\bibinfo{journal}{Science Advances}}
  \textbf{\bibinfo{volume}{3}} (\bibinfo{year}{2017}).

\bibitem{Pugh_1953}
\bibinfo{author}{Pugh, E.~M.} \& \bibinfo{author}{Rostoker, N.}
\newblock \bibinfo{title}{{Hall effect in ferromagnetic materials}}.
\newblock \emph{\bibinfo{journal}{Reviews of Modern Physics}}
  \textbf{\bibinfo{volume}{25}}, \bibinfo{pages}{151--157}
  (\bibinfo{year}{1953}).

\bibitem{nagaosa2010anomalous}
\bibinfo{author}{Nagaosa, N.}, \bibinfo{author}{Sinova, J.},
  \bibinfo{author}{Onoda, S.}, \bibinfo{author}{MacDonald, A.~H.} \&
  \bibinfo{author}{Ong, N.~P.}
\newblock \bibinfo{title}{{Anomalous Hall effect}}.
\newblock \emph{\bibinfo{journal}{Reviews of Modern Physics}}
  \textbf{\bibinfo{volume}{82}}, \bibinfo{pages}{1539--1592}
  (\bibinfo{year}{2010}).

\bibitem{Nerst_1887}
\bibinfo{author}{Nernst, W.}
\newblock \bibinfo{title}{{Ueber die electromotorischen Kr{\"{a}}fte, welche
  durch den Magnetismus in von einem W{\"{a}}rmestrome durchflossenen
  Metallplatten geweckt werden}}.
\newblock \emph{\bibinfo{journal}{Annalen der Physik}}
  \textbf{\bibinfo{volume}{267}}, \bibinfo{pages}{760--789}
  (\bibinfo{year}{1887}).

\bibitem{XiaoDi_2006}
\bibinfo{author}{Xiao, D.}, \bibinfo{author}{Yao, Y.}, \bibinfo{author}{Fang,
  Z.} \& \bibinfo{author}{Niu, Q.}
\newblock \bibinfo{title}{{Berry-phase effect in anomalous thermoelectric
  transport}}.
\newblock \emph{\bibinfo{journal}{Physical Review Letters}}
  \textbf{\bibinfo{volume}{97}}, \bibinfo{pages}{026603}
  (\bibinfo{year}{2006}).

\bibitem{haldane1988qahe}
\bibinfo{author}{Haldane, F. D.~M.}
\newblock \bibinfo{title}{Model for a quantum hall effect without landau
  levels: Condensed-matter realization of the "parity anomaly"}.
\newblock \emph{\bibinfo{journal}{Phys. Rev. Lett.}}
  \textbf{\bibinfo{volume}{61}}, \bibinfo{pages}{2015--2018}
  (\bibinfo{year}{1988}).

\bibitem{Chang2013}
\bibinfo{author}{Chang, C.-Z.} \emph{et~al.}
\newblock \bibinfo{title}{Experimental observation of the quantum anomalous
  hall effect in a magnetic topological insulator}.
\newblock \emph{\bibinfo{journal}{Science}} \textbf{\bibinfo{volume}{340}},
  \bibinfo{pages}{167--170} (\bibinfo{year}{2013}).

\bibitem{xu2011chern}
\bibinfo{author}{Xu, G.}, \bibinfo{author}{Weng, H.}, \bibinfo{author}{Wang,
  Z.}, \bibinfo{author}{Dai, X.} \& \bibinfo{author}{Fang, Z.}
\newblock \bibinfo{title}{Chern semimetal and the quantized anomalous hall
  effect in hgcr2se4}.
\newblock \emph{\bibinfo{journal}{Physical review letters}}
  \textbf{\bibinfo{volume}{107}}, \bibinfo{pages}{186806}
  (\bibinfo{year}{2011}).

\bibitem{liu2017quantum}
\bibinfo{author}{Liu, W.~E.}, \bibinfo{author}{Hankiewicz, E.~M.},
  \bibinfo{author}{Culcer, D.} \emph{et~al.}
\newblock \bibinfo{title}{Quantum transport in weyl semimetal thin films in the
  presence of spin-orbit coupled impurities}.
\newblock \emph{\bibinfo{journal}{Physical Review B}}
  \textbf{\bibinfo{volume}{96}}, \bibinfo{pages}{045307}
  (\bibinfo{year}{2017}).

\bibitem{muechler2017realization}
\bibinfo{author}{Muechler, L.}, \bibinfo{author}{Liu, E.}, \bibinfo{author}{Xu,
  Q.}, \bibinfo{author}{Felser, C.} \& \bibinfo{author}{Sun, Y.}
\newblock \bibinfo{title}{Realization of quantum anomalous hall effect from a
  magnetic weyl semimetal}.
\newblock \emph{\bibinfo{journal}{arXiv preprint arXiv:1712.08115}}
  (\bibinfo{year}{2017}).

\bibitem{sarma2006topological}
\bibinfo{author}{Sarma, S.~D.}, \bibinfo{author}{Freedman, M.} \&
  \bibinfo{author}{Nayak, C.}
\newblock \bibinfo{title}{Topological quantum computation}.
\newblock \emph{\bibinfo{journal}{Physics Today}}
  \textbf{\bibinfo{volume}{59}}, \bibinfo{pages}{32--38}
  (\bibinfo{year}{2006}).

\bibitem{heusler1903ubermagnetische}
\bibinfo{author}{Heusler, F.}
\newblock \bibinfo{title}{{{\"{U}}ber magnetische Manganlegierungen}}.
\newblock \emph{\bibinfo{journal}{Verhandlungen der Deutschen Physikalischen
  Gesellschaft}} \textbf{\bibinfo{volume}{5}}, \bibinfo{pages}{219}
  (\bibinfo{year}{1903}).

\bibitem{chadov2010}
\bibinfo{author}{Chadov, S.} \emph{et~al.}
\newblock \bibinfo{title}{{Tunable multifunctional topological insulators in
  ternary Heusler compounds}}.
\newblock \emph{\bibinfo{journal}{Nature Materials}}
  \textbf{\bibinfo{volume}{9}}, \bibinfo{pages}{541--545}
  (\bibinfo{year}{2010}).

\bibitem{wollmann2017}
\bibinfo{author}{Wollmann, L.}, \bibinfo{author}{Nayak, A.~K.},
  \bibinfo{author}{Parkin, S. S.~P.} \& \bibinfo{author}{Felser, C.}
\newblock \bibinfo{title}{{Heusler 4.0: Tunable Materials}}.
\newblock \emph{\bibinfo{journal}{Annual Review of Materials Research}}
  \textbf{\bibinfo{volume}{47}}, \bibinfo{pages}{247--270}
  (\bibinfo{year}{2016}).

\bibitem{manna2017colossal}
\bibinfo{author}{Manna, K.} \emph{et~al.}
\newblock \bibinfo{title}{{From Colossal to Zero: Controlling the Anomalous
  Hall Effect in Magnetic Heusler Compounds via Berry Curvature Design}}.
\newblock \emph{\bibinfo{journal}{Physical Review X}}
  \textbf{\bibinfo{volume}{8}} (\bibinfo{year}{2018}).

\bibitem{Manna2018}
\bibinfo{author}{Manna, K.}, \bibinfo{author}{Sun, Y.},
  \bibinfo{author}{Muechler, L.}, \bibinfo{author}{K{\"{u}}bler, J.} \&
  \bibinfo{author}{Felser, C.}
\newblock \bibinfo{title}{{Heusler, Weyl and Berry}}.
\newblock \emph{\bibinfo{journal}{Nature Reviews Materials}}
  (\bibinfo{year}{2018}).

\bibitem{glas2013anomalous}
\bibinfo{author}{Glas, M.}, \bibinfo{author}{Ebke, D.}, \bibinfo{author}{Imort,
  I.-M.}, \bibinfo{author}{Thomas, P.} \& \bibinfo{author}{Reiss, G.}
\newblock \bibinfo{title}{Anomalous hall effect in perpendicularly magnetized
  mn3- xga thin films}.
\newblock \emph{\bibinfo{journal}{Journal of magnetism and magnetic materials}}
  \textbf{\bibinfo{volume}{333}}, \bibinfo{pages}{134--137}
  (\bibinfo{year}{2013}).

\bibitem{hu2018anomalous}
\bibinfo{author}{Hu, J.} \emph{et~al.}
\newblock \bibinfo{title}{Anomalous hall and nernst effects in co 2 ti sn and
  co 2 ti 0.6 v 0.4 sn heusler thin films}.
\newblock \emph{\bibinfo{journal}{Physical Review Applied}}
  \textbf{\bibinfo{volume}{10}}, \bibinfo{pages}{044037}
  (\bibinfo{year}{2018}).

\bibitem{reichlova2018large}
\bibinfo{author}{Reichlova, H.} \emph{et~al.}
\newblock \bibinfo{title}{{Large anomalous Nernst effect in thin films of the
  Weyl semimetal Co2MnGa}}.
\newblock \emph{\bibinfo{journal}{Applied Physics Letters}}
  \textbf{\bibinfo{volume}{113}} (\bibinfo{year}{2018}).

\bibitem{markou2019thickness}
\bibinfo{author}{Markou, A.} \emph{et~al.}
\newblock \bibinfo{title}{Thickness dependence of the anomalous hall effect in
  thin films of the topological semimetal co 2 mnga}.
\newblock \emph{\bibinfo{journal}{Physical Review B}}
  \textbf{\bibinfo{volume}{100}}, \bibinfo{pages}{054422}
  (\bibinfo{year}{2019}).

\bibitem{Belopolski1278}
\bibinfo{author}{Belopolski, I.} \emph{et~al.}
\newblock \bibinfo{title}{Discovery of topological weyl fermion lines and
  drumhead surface states in a room temperature magnet}.
\newblock \emph{\bibinfo{journal}{Science}} \textbf{\bibinfo{volume}{365}},
  \bibinfo{pages}{1278--1281} (\bibinfo{year}{2019}).

\bibitem{ludbrook2017perpendicular}
\bibinfo{author}{Ludbrook, B.~M.}, \bibinfo{author}{Ruck, B.~J.} \&
  \bibinfo{author}{Granville, S.}
\newblock \bibinfo{title}{{Perpendicular magnetic anisotropy in Co2MnGa and its
  anomalous Hall effect}}.
\newblock \emph{\bibinfo{journal}{Applied Physics Letters}}
  \textbf{\bibinfo{volume}{110}}, \bibinfo{pages}{62408}
  (\bibinfo{year}{2017}).

\bibitem{Guin2018}
\bibinfo{author}{Guin, S.~N.} \emph{et~al.}
\newblock \bibinfo{title}{{Anomalous Nernst effect beyond the magnetization
  scaling relation in the ferromagnetic Heusler compound Co2MnGa}}.
\newblock \emph{\bibinfo{journal}{NPG Asia Materials}}
  \textbf{\bibinfo{volume}{11}} (\bibinfo{year}{2019}).

\bibitem{Sakai2018}
\bibinfo{author}{Sakai, A.} \emph{et~al.}
\newblock \bibinfo{title}{{Giant anomalous Nernst effect and quantum-critical
  scaling in a ferromagnetic semimetal}}.
\newblock \emph{\bibinfo{journal}{Nature Physics}}
  \textbf{\bibinfo{volume}{14}}, \bibinfo{pages}{1119--1124}
  (\bibinfo{year}{2018}).

\bibitem{kerr1877}
\bibinfo{author}{Kerr, J.}
\newblock \bibinfo{title}{{On rotation of the plane of polarization by
  reflection from the pole of a magnet}}.
\newblock \emph{\bibinfo{journal}{The London, Edinburgh, and Dublin
  Philosophical Magazine and Journal of Science}} \textbf{\bibinfo{volume}{3}},
  \bibinfo{pages}{321--343} (\bibinfo{year}{1877}).

\bibitem{PhysRevB.75.214416}
\bibinfo{author}{Kim, M.~H.} \emph{et~al.}
\newblock \bibinfo{title}{{Determination of the infrared complex
  magnetoconductivity tensor in itinerant ferromagnets from Faraday and Kerr
  measurements}}.
\newblock \emph{\bibinfo{journal}{Physical Review B}}
  \textbf{\bibinfo{volume}{75}}, \bibinfo{pages}{214416}
  (\bibinfo{year}{2007}).

\bibitem{kahn1969ultraviolet}
\bibinfo{author}{Kahn, F.~J.}, \bibinfo{author}{Pershan, P.~S.} \&
  \bibinfo{author}{Remeika, J.~P.}
\newblock \bibinfo{title}{{Ultraviolet magneto-optical properties of
  single-crystal orthoferrites, garnets, and other ferric oxide compounds}}.
\newblock \emph{\bibinfo{journal}{Physical Review}}
  \textbf{\bibinfo{volume}{186}}, \bibinfo{pages}{891--918}
  (\bibinfo{year}{1969}).

\bibitem{Thouless_1982}
\bibinfo{author}{Thouless, D.~J.}, \bibinfo{author}{Kohmoto, M.},
  \bibinfo{author}{Nightingale, M.~P.} \& \bibinfo{author}{den Nijs, M.}
\newblock \bibinfo{title}{{Quantized Hall Conductance in a Two-Dimensional
  Periodic Potential}}.
\newblock \emph{\bibinfo{journal}{Phys. Rev. Lett.}}
  \textbf{\bibinfo{volume}{49}}, \bibinfo{pages}{405--408}
  (\bibinfo{year}{1982}).

\bibitem{Xiao2010}
\bibinfo{author}{Xiao, D.}, \bibinfo{author}{Chang, M.-C.} \&
  \bibinfo{author}{Niu, Q.}
\newblock \bibinfo{title}{{Berry Phase Effects on Electronic Properties}}.
\newblock \emph{\bibinfo{journal}{Rev. Mod. Phys.}}
  \textbf{\bibinfo{volume}{82}}, \bibinfo{pages}{1--48} (\bibinfo{year}{2009}).

\bibitem{avron2003topological}
\bibinfo{author}{Avron, J.~E.}, \bibinfo{author}{Osadchy, D.} \&
  \bibinfo{author}{Seiler, R.}
\newblock \bibinfo{title}{A topological look at the quantum hall effect}.
\newblock \emph{\bibinfo{journal}{Physics today}}
  \textbf{\bibinfo{volume}{56}}, \bibinfo{pages}{38--42}
  (\bibinfo{year}{2003}).

\bibitem{alabama}
\bibinfo{title}{{Heusler Database}} (\bibinfo{year}{2015}).
\newblock \urlprefix\url{http://heusleralloys.mint.ua.edu/}.

\bibitem{PhysRevB.96.235103}
\bibinfo{author}{Rauch, T. c.~v.}, \bibinfo{author}{Nguyen~Minh, H.},
  \bibinfo{author}{Henk, J.} \& \bibinfo{author}{Mertig, I.}
\newblock \bibinfo{title}{Model for ferromagnetic weyl and nodal line
  semimetals: Topological invariants, surface states, anomalous and spin hall
  effect}.
\newblock \emph{\bibinfo{journal}{Phys. Rev. B}} \textbf{\bibinfo{volume}{96}},
  \bibinfo{pages}{235103} (\bibinfo{year}{2017}).

\bibitem{noky2018characterization}
\bibinfo{author}{Noky, J.}, \bibinfo{author}{Gooth, J.},
  \bibinfo{author}{Felser, C.} \& \bibinfo{author}{Sun, Y.}
\newblock \bibinfo{title}{{Characterization of topological band structures away
  from the Fermi level by anomalous Nernst measurements}}.
\newblock \emph{\bibinfo{journal}{Physical Review B}}
  \textbf{\bibinfo{volume}{98}}, \bibinfo{pages}{1--5} (\bibinfo{year}{2018}).

\bibitem{noky2018large}
\bibinfo{author}{Noky, J.}, \bibinfo{author}{Xu, Q.}, \bibinfo{author}{Felser,
  C.} \& \bibinfo{author}{Sun, Y.}
\newblock \bibinfo{title}{{Large anomalous Hall and Nernst effects from nodal
  line symmetry breaking in Fe2MnX (X = P, As, Sb)}}.
\newblock \emph{\bibinfo{journal}{Phys. Rev. B}} \textbf{\bibinfo{volume}{99}},
  \bibinfo{pages}{165117} (\bibinfo{year}{2019}).

\bibitem{kim2018large}
\bibinfo{author}{Kim, K.} \emph{et~al.}
\newblock \bibinfo{title}{{Large anomalous Hall current induced by topological
  nodal lines in a ferromagnetic van der Waals semimetal}}.
\newblock \emph{\bibinfo{journal}{Nature Materials}}
  \textbf{\bibinfo{volume}{17}}, \bibinfo{pages}{794--799}
  (\bibinfo{year}{2018}).

\bibitem{miyasato2007crossover}
\bibinfo{author}{Miyasato, T.} \emph{et~al.}
\newblock \bibinfo{title}{{Crossover behavior of the anomalous hall effect and
  anomalous nernst effect in itinerant ferromagnets}}.
\newblock \emph{\bibinfo{journal}{Physical Review Letters}}
  \textbf{\bibinfo{volume}{99}}, \bibinfo{pages}{1--4} (\bibinfo{year}{2007}).

\bibitem{WEBSTER19711221}
\bibinfo{author}{Webster, P.~J.}
\newblock \bibinfo{title}{{Magnetic and chemical order in Heusler alloys
  containing cobalt and manganese}}.
\newblock \emph{\bibinfo{journal}{Journal of Physics and Chemistry of Solids}}
  \textbf{\bibinfo{volume}{32}}, \bibinfo{pages}{1221--1231}
  (\bibinfo{year}{1971}).

\bibitem{masumoto1972new}
\bibinfo{author}{Masumoto, H.} \& \bibinfo{author}{Watanabe, K.}
\newblock \bibinfo{title}{{New Compounds of the Clb, Cl Types of RhMnSb, IrMnSn
  and IrMnAl, New L21 (Heusler) Type of Ir2MnAl and Rh2MnAl Alloys, and
  Magnetic Properties}}.
\newblock \emph{\bibinfo{journal}{Journal of the Physical Society of Japan}}
  \textbf{\bibinfo{volume}{32}}, \bibinfo{pages}{281} (\bibinfo{year}{1971}).

\bibitem{SUITS1976423}
\bibinfo{author}{Suits, J.~C.}
\newblock \bibinfo{title}{{Structural instability in new magnetic heusler
  compounds}}.
\newblock \emph{\bibinfo{journal}{Solid State Communications}}
  \textbf{\bibinfo{volume}{18}}, \bibinfo{pages}{423--425}
  (\bibinfo{year}{1976}).

\bibitem{kresse1996}
\bibinfo{author}{Kresse, G.} \& \bibinfo{author}{Furthm{\"{u}}ller, J.}
\newblock \bibinfo{title}{{Efficiency of ab-initio total energy calculations
  for metals and semiconductors using a plane-wave basis set}}.
\newblock \emph{\bibinfo{journal}{Computational Materials Science}}
  \textbf{\bibinfo{volume}{6}}, \bibinfo{pages}{15--50} (\bibinfo{year}{1996}).

\bibitem{perdew1996}
\bibinfo{author}{Perdew, J.~P.}, \bibinfo{author}{Burke, K.} \&
  \bibinfo{author}{Ernzerhof, M.}
\newblock \bibinfo{title}{{Generalized gradient approximation made simple}}.
\newblock \emph{\bibinfo{journal}{Physical Review Letters}}
  \textbf{\bibinfo{volume}{77}}, \bibinfo{pages}{3865--3868}
  (\bibinfo{year}{1996}).

\bibitem{Mostofi2008}
\bibinfo{author}{Mostofi, A.~A.} \emph{et~al.}
\newblock \bibinfo{title}{{wannier90: A tool for obtaining maximally-localised
  Wannier functions}}.
\newblock \emph{\bibinfo{journal}{Computer Physics Communications}}
  \textbf{\bibinfo{volume}{178}}, \bibinfo{pages}{685--699}
  (\bibinfo{year}{2008}).

\bibitem{PhysRevLett.92.037204}
\bibinfo{author}{Yao, Y.} \emph{et~al.}
\newblock \bibinfo{title}{{First Principles Calculation of Anomalous Hall
  Conductivity in Ferromagnetic bcc Fe}}.
\newblock \emph{\bibinfo{journal}{Physical Review Letters}}
  \textbf{\bibinfo{volume}{92}}, \bibinfo{pages}{4} (\bibinfo{year}{2004}).

\bibitem{ebert1996magneto}
\bibinfo{author}{Chemie, U.}
\newblock \bibinfo{title}{{Magneto-optical effects in transition metal
  systems}}.
\newblock \emph{\bibinfo{journal}{Reports on Progress in Physics}}
  \textbf{\bibinfo{volume}{59}}, \bibinfo{pages}{1665--1735}
  (\bibinfo{year}{1996}).

\end{thebibliography}

\end{document}